\documentclass[12pt]{article}
\usepackage{graphicx} 
\begin{document}
{\bf Simulation of majority rule disturbed by power-law noise on directed and undirected Barab\'asi-Albert networks }

\bigskip
\centerline{F.W.S. Lima }
 
\bigskip
\noindent
Departamento de F\'{\i}sica, 
Universidade Federal do Piau\'{\i}, 64049-550, Teresina - PI, Brazil \\

\medskip
  e-mail:  wel@ufpi.br
\bigskip
 
{\small Abstract: On {\it directed} and {\it undirected} Barab\'asi-Albert networks the
 Ising model with spin $S=1/2$ in the presence of a kind of noise is now studied through Monte Carlo   simulations. The noise spectrum $P(n)$ follows a power law, where $P(n)$ is the probability of flipping randomly select $n$ spins at each time step. The noise spectrum $P(n)$ is introduced to mimic the self-organized criticality as a model influence of a complex environment.
 In this model, different from the square lattice, the order-disorder phase transition of the
 order parameter not is observed. For {\it directed}  Barab\'asi-Albert networks the magnetisation tends to zero exponentially and {\it undirected} Barab\'asi-Albert networks , its remain constant.
}
 
 
\bigskip

 {\bf Introduction}

 This paper deals with Ising spin on {\it directed} and {\it undirected}     Barab\'asi-Albert(BA) networks in the presence of a noise. Sumour and Shabat \cite{sumour,sumourss} investigated Ising models with
 spin $S=1/2$ on {\it directed} BA networks \cite{ba} with
 the usual Glauber dynamics.  No spontaneous magnetisation was 
 found, in contrast to the case of {\it undirected}  BA networks
 \cite{alex,indekeu,bianconi} where a spontaneous magnetisation was
 found below a critical temperature which increases logarithmically with
 system size. Lima and Stauffer \cite{lima} simulated
 {\it directed} square, cubic and hypercubic lattices in two to five    dimensions
 with heat bath dynamics in order to separate the network effects of directedness. They also compared different spin flip
 algorithms, including cluster flips \cite{wang}, for
 Ising-BA networks. They found a freezing-in of the 
 magnetisation similar to  \cite{sumour,sumourss}, following an Arrhenius
 law at least in low dimensions. This lack of a spontaneous magnetisation
 (in the usual sense)
 is consistent with the fact
 that if on a directed lattice a spin $S_j$ influences spin $S_i$, then
 spin $S_i$ in turn does not influence $S_j$, 
 and there may be no well-defined total energy. Thus, they show that for
 the same  scale-free networks, different algorithms give different
 results. Recently Stauffer and Ku{\l}akowski \cite{DK} simulated the Ising two-dimensional ferromagnet in the presence of a special kind of noise different from the traditional temperature and Boltzmann probabilities, where the noise consist of at each iteration of a $L \times L$ square lattice with four neighbours for each sites, besides the above majority rule they select $n$ times randomly a spin and flip it. The probability distribution function $P(n)$ of these numbers $n$ is taken as a power law,
\begin{equation}
%\begin{center}
P(n)\propto 1/n^{\alpha}.
%\end{center}
\end{equation}

\begin{figure}[hbt]
\begin{center}
\includegraphics [angle=-90,scale=0.5]{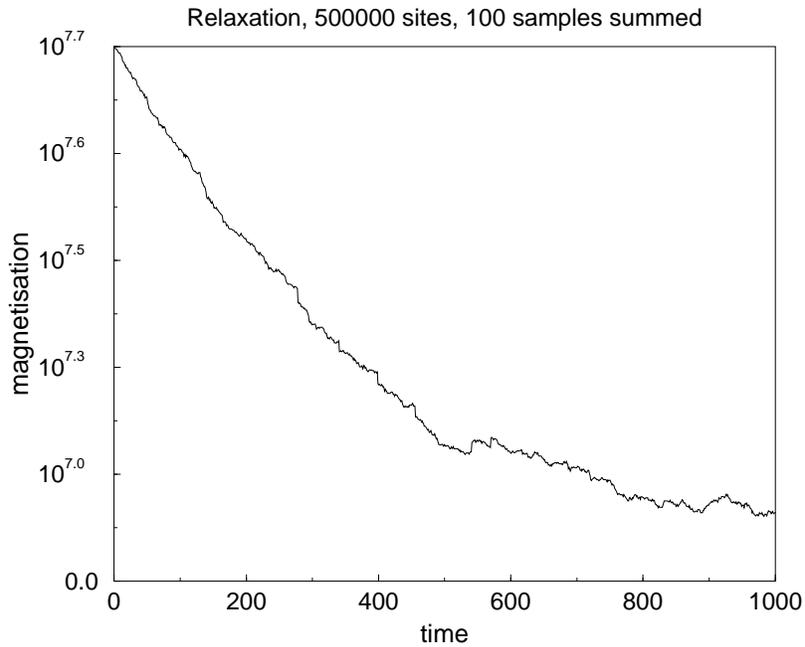}
\end{center}
\caption{Summed magnetisations versus time, $T=1$, $\alpha=1$, and $m=2$ for {\it directed} (BA) networks  .}
\end{figure}
\bigskip

\begin{figure}[hbt]
\begin{center}
\includegraphics [angle=-90,scale=0.29]{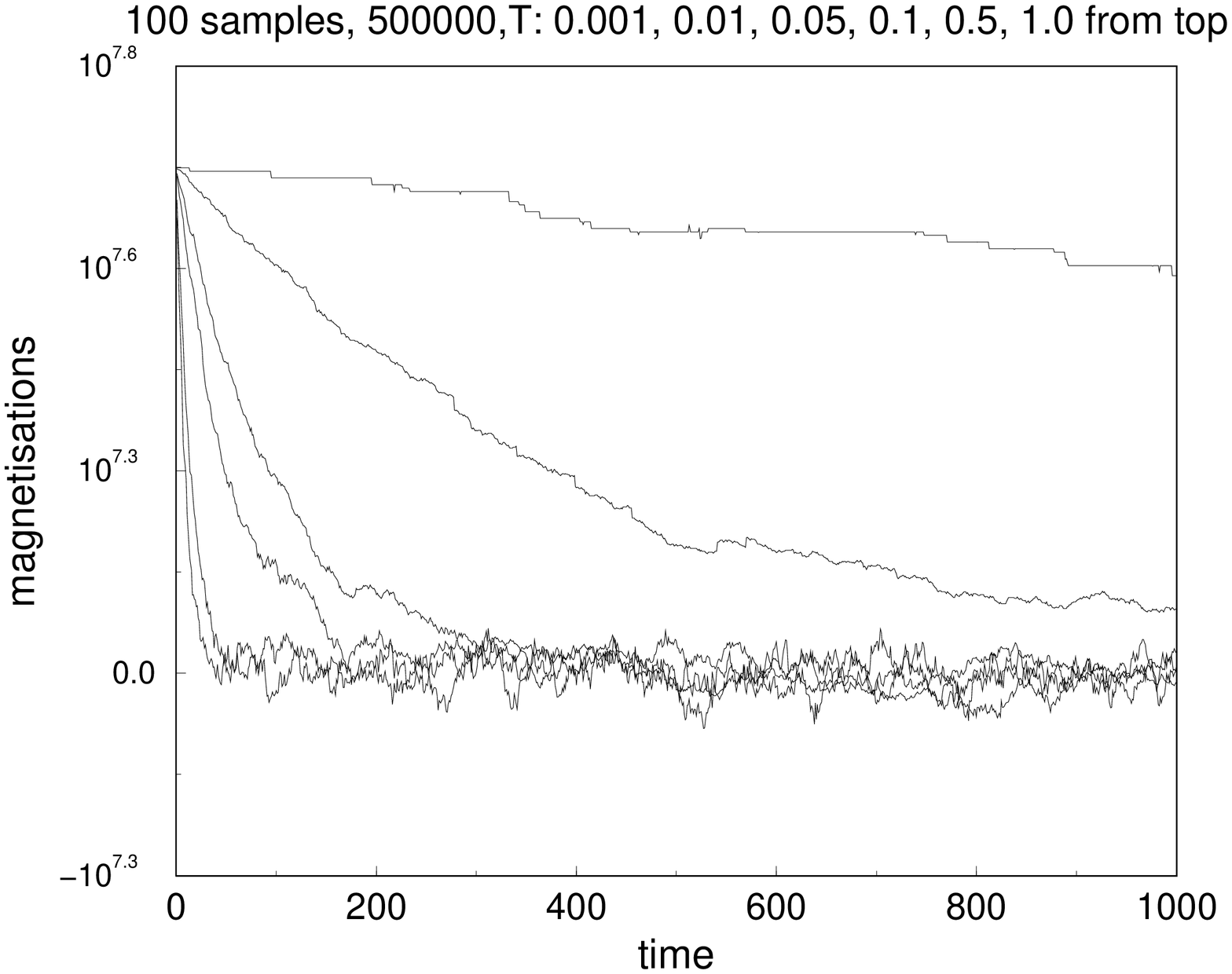}
\includegraphics[angle=-90,scale=0.29]{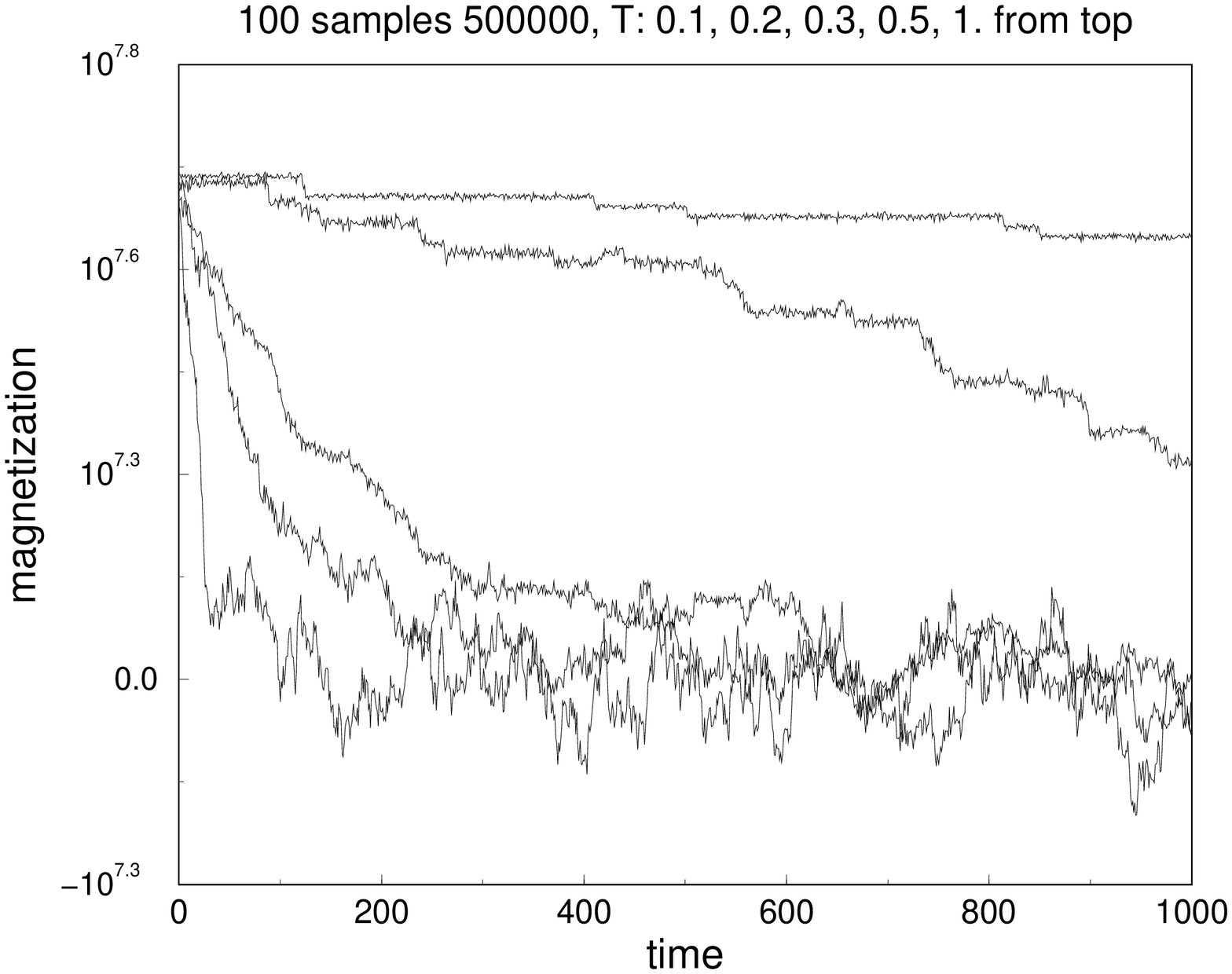}
\end{center}
\caption{Summed magnetisations versus time for $0.01\le T \le 1.0$, $\alpha=1$ and $m=2$ (left) and $0.1 \le T \le 1.0$, $\alpha=1$ and $m=7$ (right) for {\it directed} (BA) networks.  }
\end{figure} 
\bigskip

\begin{figure}[hbt]
\begin{center}
\includegraphics[angle=-90,scale=0.29]{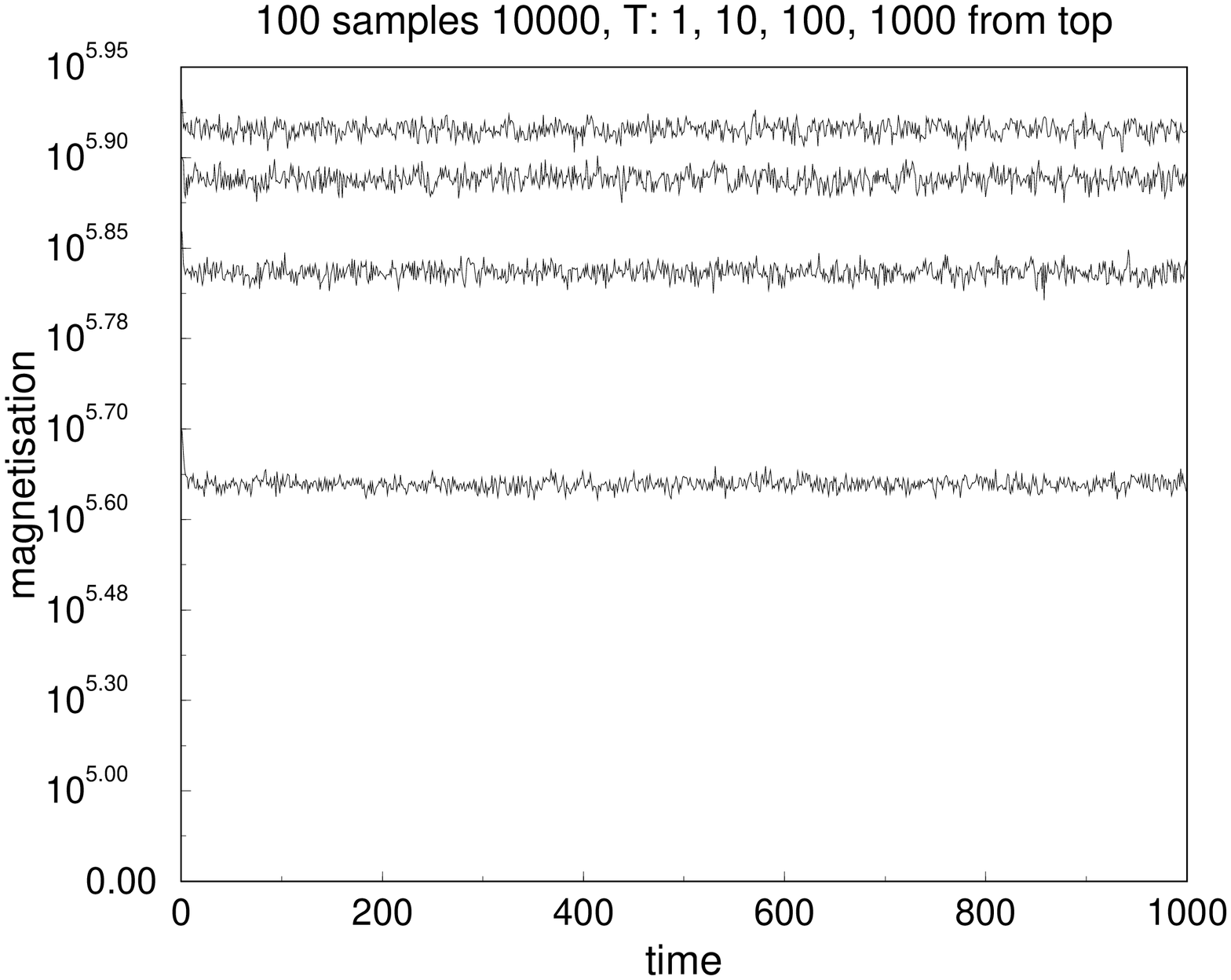}
\includegraphics[angle=-90,scale=0.29]{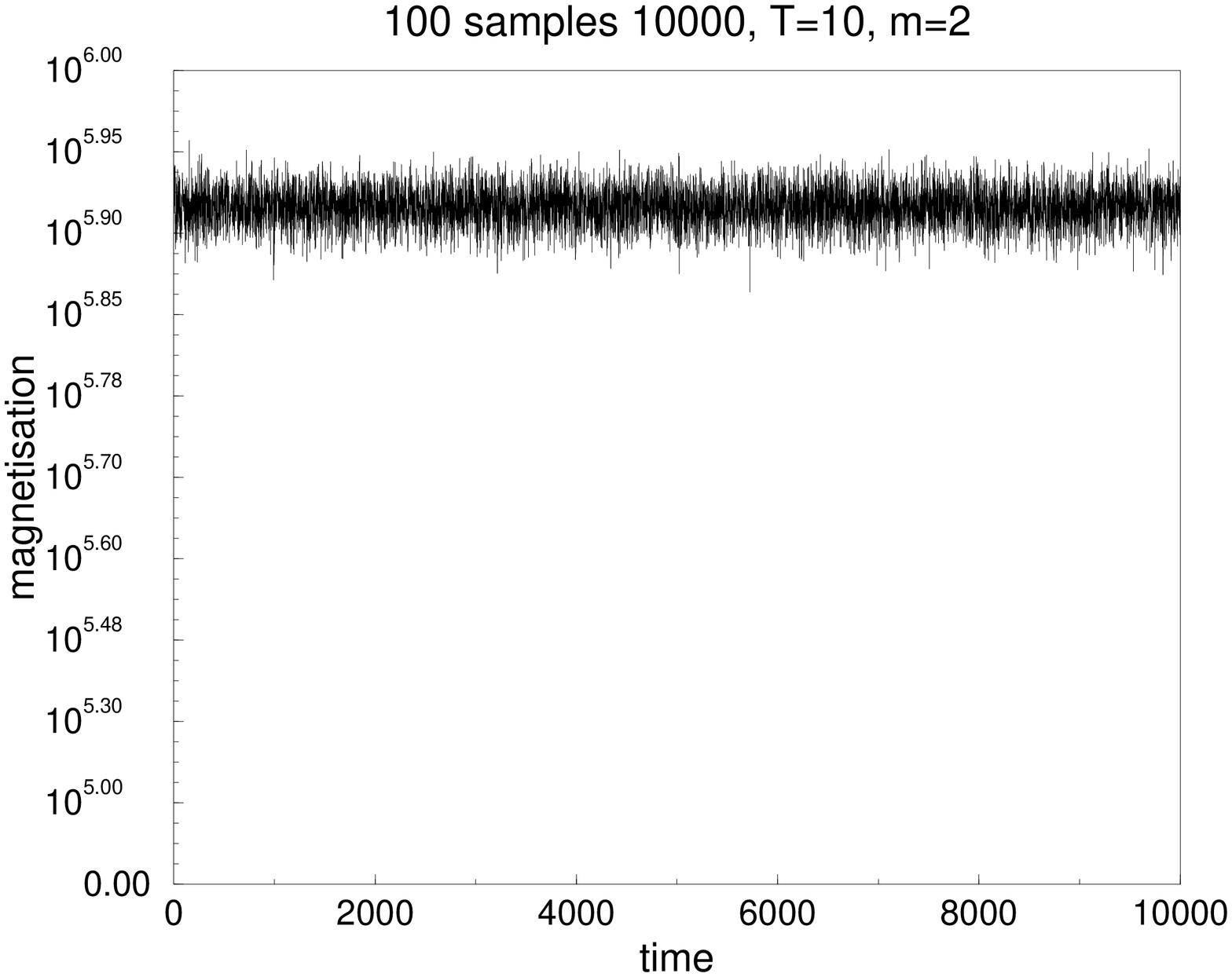}
\end{center}
\caption{
Summed magnetisations versus time for $1 \le T \le 1000.$, $\alpha=1$ and $m=2$ (left) and for $ T=10 $, $\alpha=1$ and $m=2$ (right) for {\it undirected} (BA) networks.}
\end{figure}
 
\begin{figure}[hbt]
\begin{center}
\includegraphics[angle=-90,scale=0.29]{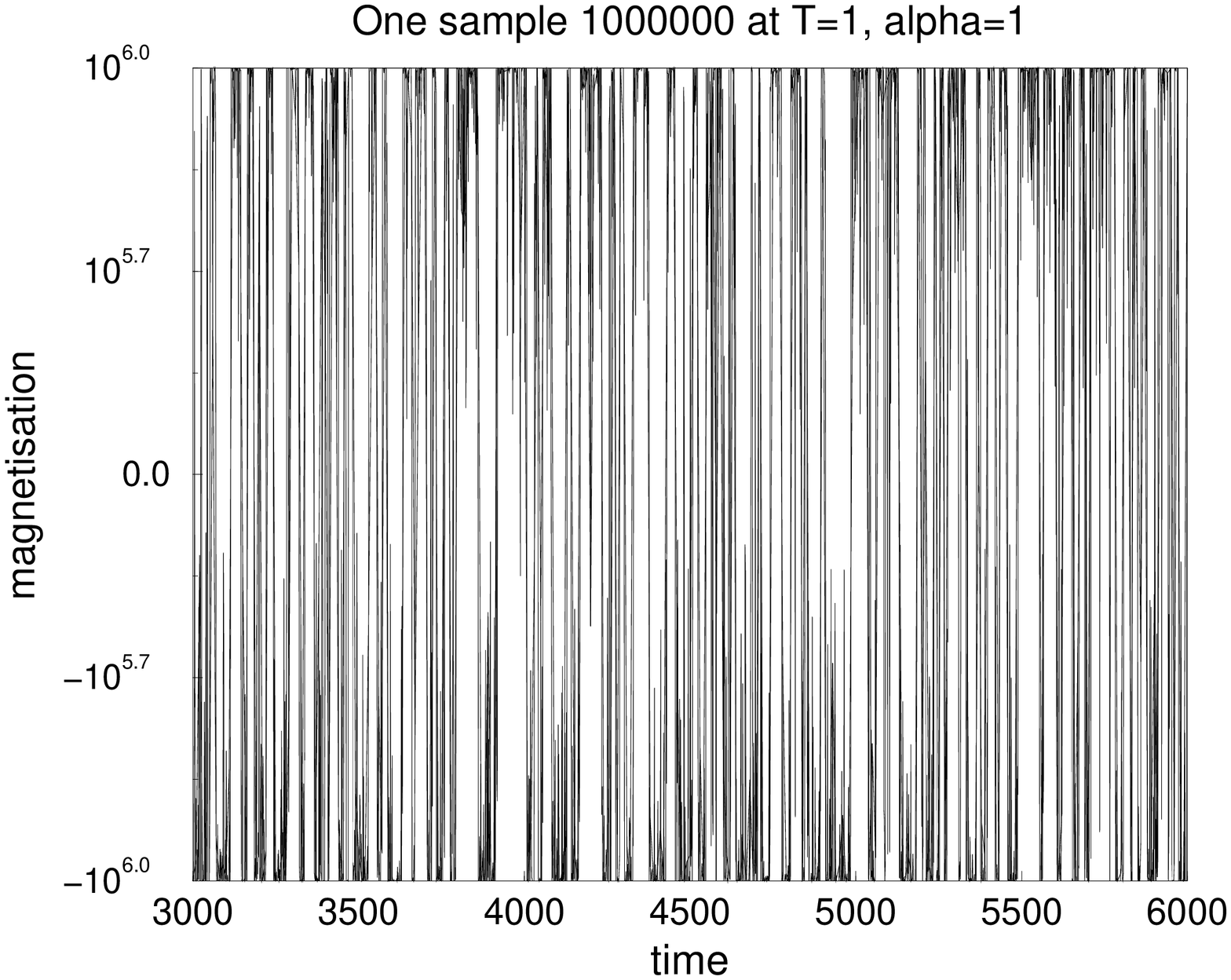}
\includegraphics[angle=-90,scale=0.29]{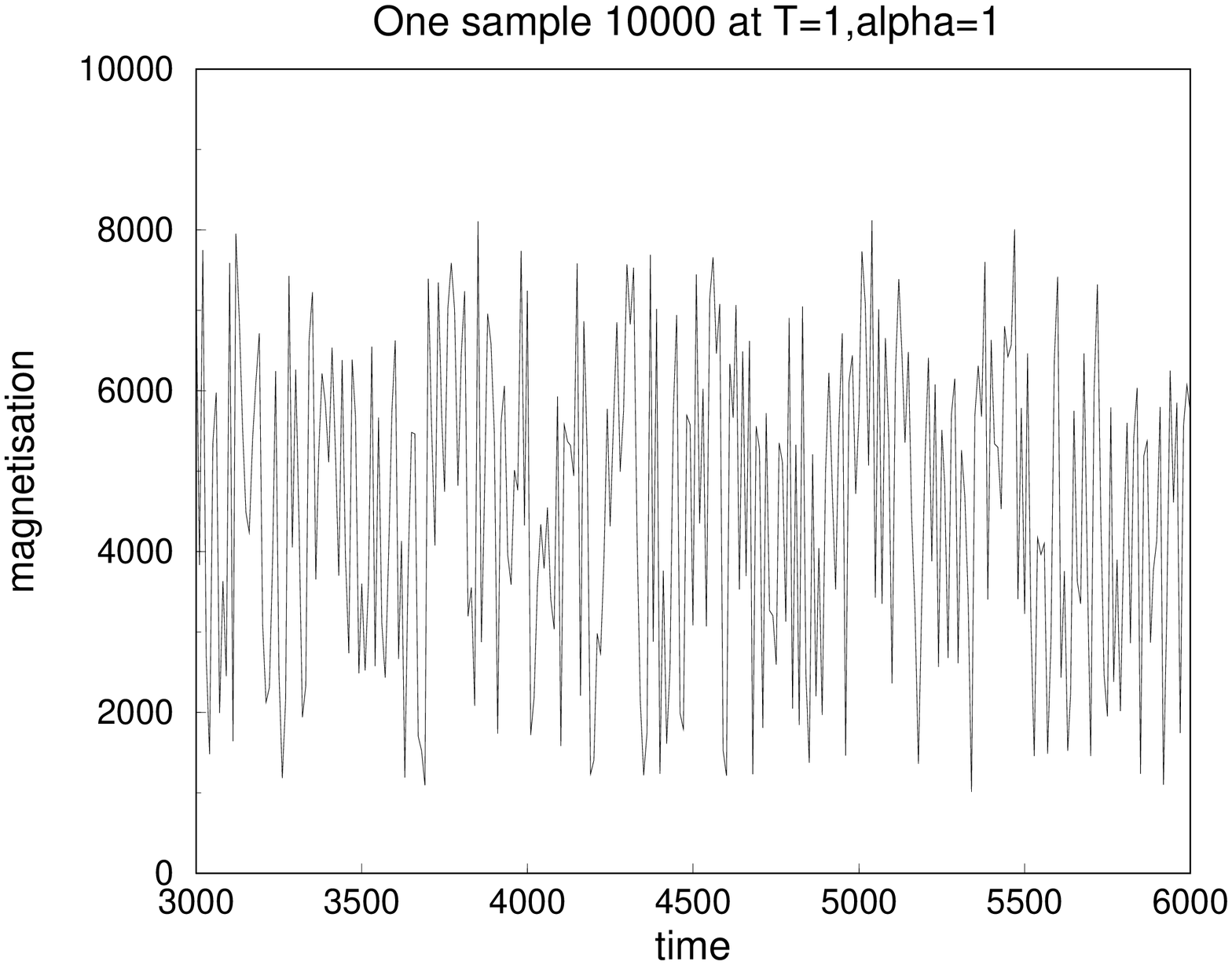}
\end{center}
\caption{Magnetization versus time for one sample, $m=2$ for {\it directed} (left) and {\it undirected}(right) (BA) networks.  } 
\end{figure}

\bigskip

\begin{figure}[hbt]
\begin{center}
\includegraphics[angle=-90,scale=0.50]{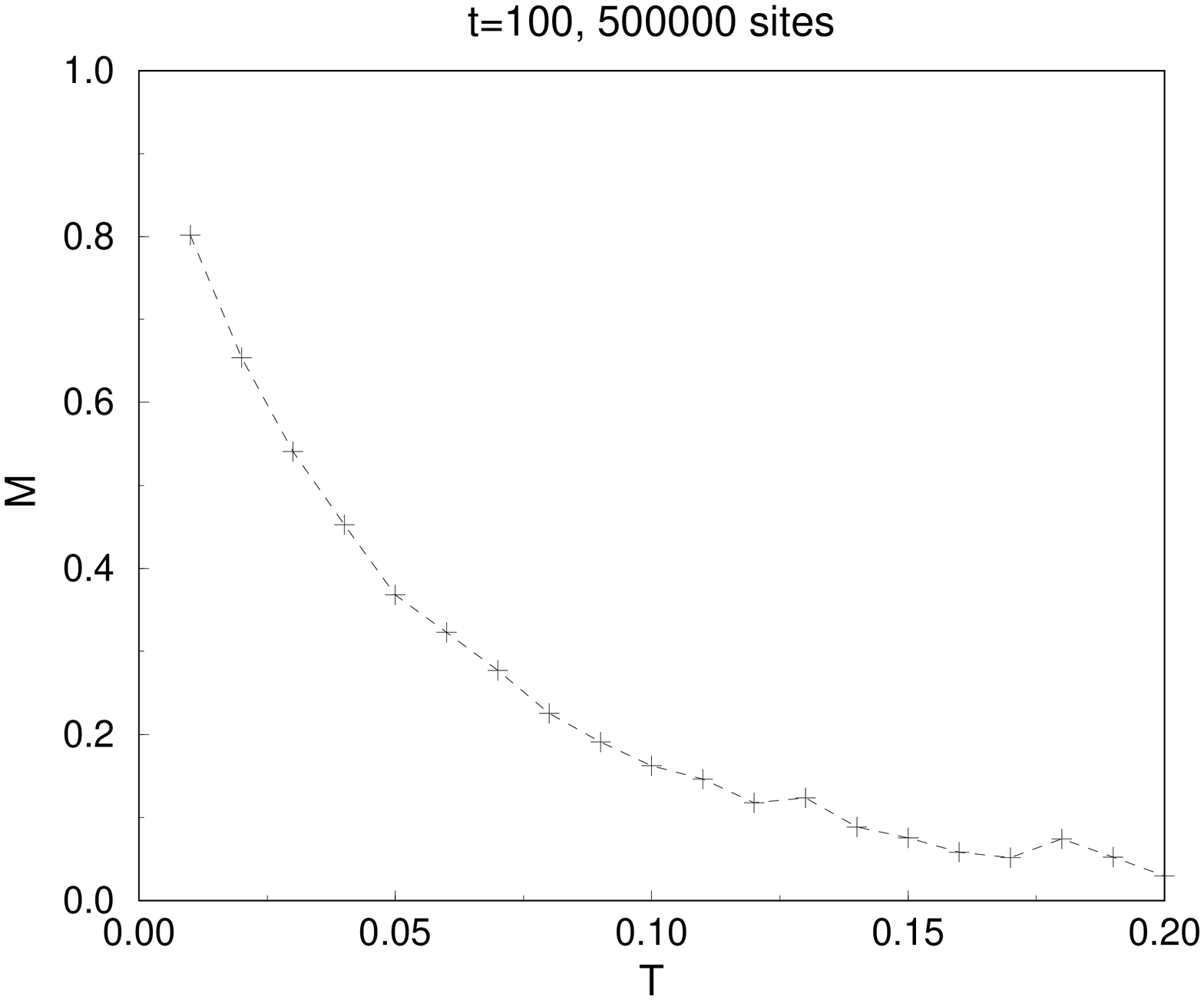}
\end{center}
\caption{Normalized magnetization, average over 1000 samples, 10 networks and
over $100< t \le 1000$, $m=2$ for {\it directed} (BA) networks.  } 
\end{figure}
\bigskip
In order to get this distribution, they determined random numbers $r$, homogeneously distributed between zero and one, and then took
\begin{equation}
%\begin{center}
n=TL^{2}r^{1/(1-\alpha)},
%\end{center}
\end{equation}
\bigskip
for $\alpha < 1$, where $1\le n \le L^{2}$ and 
\begin{equation}
%\begin{center}
n=T \exp(r \ln(L^{2}))
%\end{center}
\end{equation}
for $\alpha=1$ and $T$ determines the amplitude of the noise. Here we simulated the same model described above on {\it directed} and {\it undirected} (BA) networks and our results are different from the results obtained by Stauffer and Ku{\l}akowski\cite{DK}.

\bigskip
{\bf Results}

In Fig. 1 we show the dependence of the summed magnetisation $M=\sum_{i}S_{i}$ versus time for $\alpha=1$, $T=0.01$(eq.3) and connection number $m=2$ on {\it directed} (BA) networks with 500000 sites, when we started with all spins up. $M$ relaxes exponentially towards zero. In Fig. 2 we plot same picture of Fig.1, but for several values of $T$ and connection numbers $m=2$(left) and $m=7$(right), where for smaller  $T$ the magnetisation tries to remain positive. In the Fig.3 we plot the dependence of the summed magnetisation versus time for $\alpha=1$, several values of $T$(left), and for $T=10$(right) even up to 10000 Monte Carlo
steps with $m=2$ on {\it undirected} (BA) networks with $10000$ sites. The summed magnetisation remains positive and does not decay towards zero. It keeps fluctuating around a positive value of the magnetisation depending on the value of $T$. 

In Fig. 4 the magnetisation is shown as a function of 
time for one sample for $T=1$, $\alpha=1$ and $m=2$ on {\it directed}(left) and {\it undirected}(right)(BA) networks. For {\it directed} (BA) networks the summed magnetisation changes between positive and negative values with time, and for {\it undirected} (BA) networks the magnetisation fluctuates around a positive value.
In Fig. 5 we display the normalized magnetisation versus the temperature $T$ for {\it directed} (BA) networks. It decays exponentially towards zero for $100<t\le 1000$
which shows that a phase transition for this model does not exist, in agreement with the results for Ising model on this network \cite{sumour,sumourss}, but in disagreement with the results of Stauffer and Ku{\l}akowski\cite{DK} for square
lattice. The absolute value of the magnetisation for some  $T$ values ( no shown here) fluctuates around a positive value of the magnetisation depending on the value of $T$, in agreement with Fig. 4.

\bigskip
 
{\bf Discussion}
 
 We have presented the simulation of majority rule disturbed by power-law noise on directed and undirected (BA) networks. We study this model for case of $\alpha=1$, eq.3, and values different from 1 (not shown here), eq.2, in both cases the results are the same. For {\it directed} (BA) networks the magnetisation presents no phase transition and decays exponentially with time and with temperature. For the {\it undirected} (BA) networks the magnetisation also presents no indication of a phase transition in the presence of the noise; it assumes positive values that depend on the values of the temperature. These results are differentfrom the results obtained by Stauffer and Ku{\l}akowski\cite{DK} for a square lattice.

We acknowledge the Brazilian agency FAPEPI
(Teresina-Piau\'{\i}-Brasil) for  its financial support. This work also was supported the
system SGI Altix 1350 the computational park CENAPAD.UNICAMP-USP, SP-BRAZIL.

\end{document}